\documentclass[journal=jacsat,manuscript=article]{achemso}
\usepackage{amsmath}           % Equation module - AMS math package
\usepackage{graphicx}          % Figure module - graphicx package
\usepackage[version=3]{mhchem}
\usepackage{dcolumn}           % Align table columns on decimal point
\usepackage{xcolor}

\author{Euihwan Do}		    % Author name
\altaffiliation{Department of Physics, Pohang University of Science and Technology (POSTECH), Pohang 37673, Republic of Korea}
\author{Jae Whan Park}		    % Author name
\affiliation{Center for Artificial Low Dimensional Electronic Systems, Institute for Basic Science (IBS), Pohang 37673, Republic of Korea}
\author{Oleksandr Stetsovych}
\author{Pavel Jelinek}
\affiliation{Institute of Physics of the Czech Academy of Sciences, Cukrovarnicka 10, 18221 Prague 6, Czech Republic}
\author{Han Woong Yeom}
\email{yeom@postech.ac.kr}	    % E-mail address for corresponding author
\affiliation{Center for Artificial Low Dimensional Electronic Systems, Institute for Basic Science (IBS), Pohang 37673, Republic of Korea}
\altaffiliation{Department of Physics, Pohang University of Science and Technology (POSTECH), Pohang 37673, Republic of Korea}

\title{$Z_{3}$ Charge Density Wave of Silicon Atomic Chains on a Vicinal Silicon Surface}       % Title of the document

\abbreviations{CDW,STM/S,AFM}

\begin{document}               % Initiate the document

\begin{abstract}				% Abstract
An ideal one-dimensional electronic system is formed along atomic chains on Au-decorated vicinal silicon surfaces but the nature of its low temperature phases has been puzzled for last two decades.
Here, we unambiguously identify the low temperature structural distortion of this surface using high resolution atomic force microscopy and scanning tunneling microscopy.
The most important structural ingredient of this surface, the step-edge Si chains are found to be strongly buckled, every third atoms down, forming trimer unitcells.
This observation is consistent with the recent model of rehybridized dangling bonds and rules out the antiferromagnetic spin ordering proposed earlier.
The spectroscopy and electronic structure calculation indicate a charge density wave insulator with a $Z_{3}$ topology making it possible to exploit topological phases and excitations.
Tunneling current was found to substantially lower the energy barrier between three degenerate CDW states, which induces a dynamically fluctuating CDW at very low temperature.

\end{abstract}

\noindent{\fontfamily{phv}\selectfont \textbf{Keywords}}
topological soliton, charge density wave, atomic wire, scanning tunneling microscopy, atomic force microscopy, density functional theory

%%%%%%%%%%%%%%%%%%%%%%%%%% Introduction %%%%%%%%%%%%%%%%%%%%%%%%%%
% Nonequilibrium electronic phase

\noindent{\fontfamily{phv}\selectfont \textbf{INTRODUCTION}}

One-dimensional (1D) electronic systems have attracted great research interest due to their simplicity with exactly solvable models, the exploded manifestation of many-body interactions, and topological properties.
Non-Fermi liquid behaviors \cite{Auslaender2002,Ishii2003}, charge/spin density wave phenomena \cite{Yeom1999,Zeng2008,Schlappa2012}, nontrivial topology \cite{Cheon2015}, and topological excitations \cite{Kim2017} are among notable issues discussed actively.
While bulk compounds with strongly anisotropic structures and organic crystals \cite{Island2017} have been widely investigated materials hosting 1D electronic systems, self-assembled \cite{Do2015,Cheon2015,Kim2017,Segovia1999,Losio2001,Ahn2003,Erwin2010,Hafke2016,Braun2018,Nadj-Perge2014} or atom-manipulated \cite{Folsch2014,Drost2017} atomic chain structures on solid surfaces are in the spotlight recently.
As notable examples of self-assembled atomic chains, charge density wave (CDW) and a new type of solitons were discussed in indium chain structures on silicon surfaces \cite{Cheon2015,Kim2017}, non-Fermi liquid behaviors \cite{Segovia1999,Losio2001,Ahn2003} and quantum spin chains \cite{Erwin2010,Hafke2016,Braun2018} in gold-induced chains on silicon surfaces, and Majorana edge modes in iron chain structures on a lead surface \cite{Nadj-Perge2014}.

\begin{figure}[!tb] %[t], [b], ...
\includegraphics[width=8.6cm]{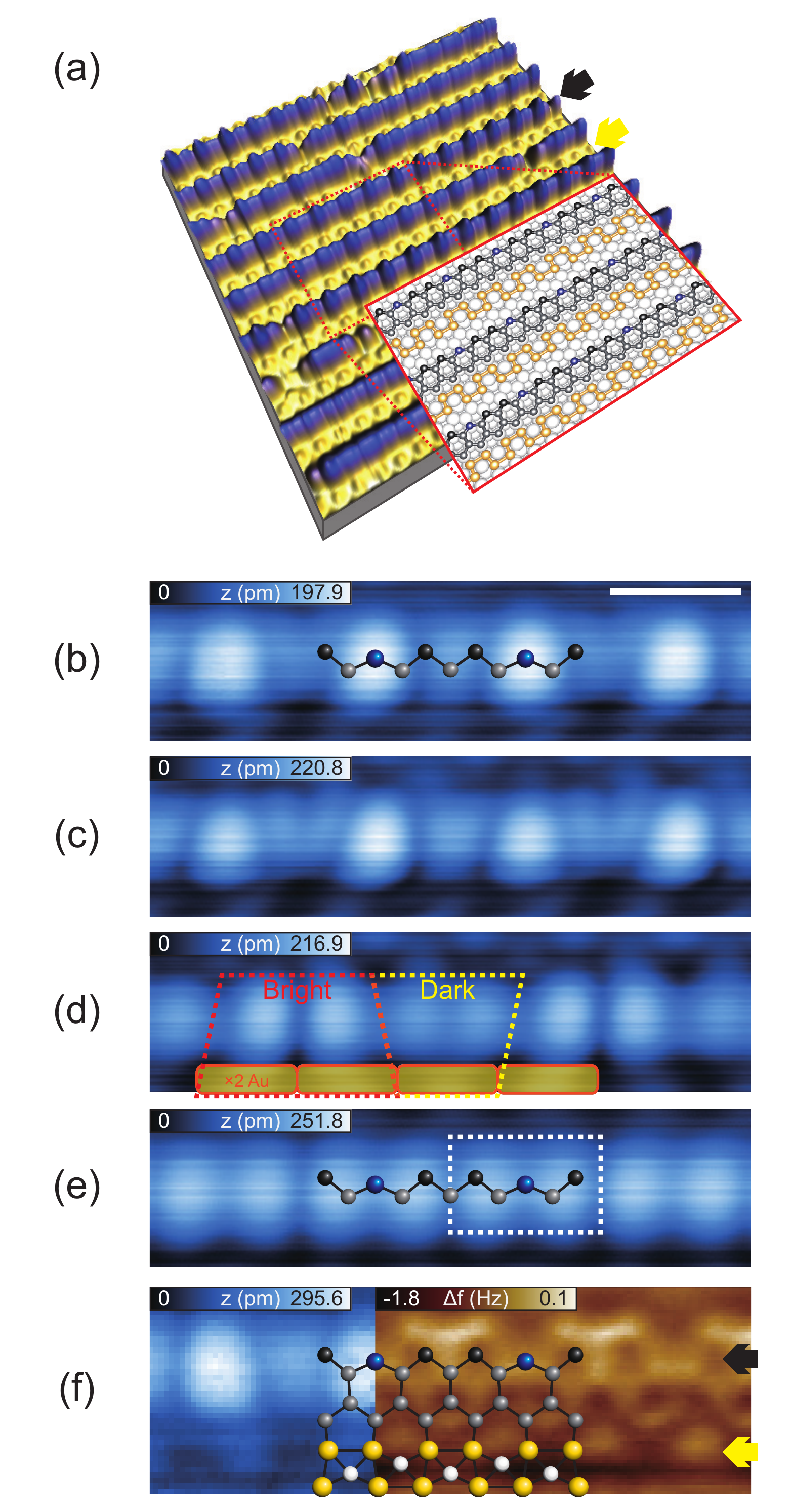}
\caption{\label{f1} (a) Topographic STM image (13.6$\times$13.6 nm$^2$ at +0.9 V sample bias) of Si(553)-Au at 50 K and schematics of the rehybridized model \cite{Braun2018}. Si (Au) atoms are represented by small (large) spheres. High resolution (b,c) empty- and (d,e) filled-state STM images (at $V$=+0.9, +0.4, -0.4, and -0.7 V, respectively, with $I$=10 pA, scale bar=1 nm) for four unit cells of the $\times$3 step-edge chain. The scale bar in (b) corresponds to 1.0 nm. (f) Comparison between 50 K STM ($V$=+0.9 V and $I$=5 pA) and 4.3 K AFM ($V$=0 V) images resolving clearly step-edge Si (black arrow) and terrace Au (yellow) chains.
}
\end{figure}

% General description for Si(553)-Au
%Here we investigate the interplay of an one-dimensional (1D) quantum phase and its transient excitation using the self-assembled atomic chain structure of the Si(553)-Au surface .
Among those representative atomic chain systems, Au-induced atomic chain structures on stepped Si surfaces have remained puzzled with intriguing phase transitions of elusive origin \cite{Crain2003,Crain2004,Segovia1999,Losio2001,Ahn2003,Ahn2005,Snijders2006,Erwin2010,Hafke2016,Braun2018,Yeom2014}.
The stepped surface of Si(553) with submonolayer Au deposited has been known to feature well ordered atomic chain arrays, which are composed basically of Au atomic rows embedded into the topmost Si layer and honeycomb Si chains formed along its step edges [Fig. \ref{f1}(a)] \cite{Krawiec2010,Erwin2010,Hafke2016,Braun2018}.
The latter provides chains of Si dangling bonds localized on step edges.
These atomic chains were discovered to exhibit two different symmetry-broken structures at low temperature; $\times$2 and $\times$3 orderings on terrace Au and step-edge Si chains, respectively [2$a_{0}$ and 3$a_{0}$ with $a_{0}$=0.384 nm, the Si lattice constant along steps].
Origins of the superstructures have been controversial for a long time \cite{Ahn2005,Snijders2006,Erwin2010,Hafke2016,Hafke2020,Braun2018,Braun2020,Edler2019,Plaickner2021}.
The recent electron diffraction study revealed that the $\times$2 structure along the terrace Au chains has a static structural distortion without a drastic temperature dependence \cite{Edler2019,Hafke2020}, which is consistently supported by the recent density functional theory (DFT) calculations \cite{Braun2018}.
However, the origin of the temperature-induced $\times$3 ordering along Si step-edge chains \cite{Ahn2005,Shin2012} is still unclear with the suggestions of the electronic/structural instability \cite{Braun2018,Braun2020,Plaickner2021} and the antiferromagnetic spin ordering on Si dangling bonds \cite{Erwin2010}.
Based on the latter, even a quantum spin liquid phase was suggested \cite{Hafke2016,Hafke2020}.
Namely, there exist two lines of models, which are distinctly based on charge/structural and spin degrees of freedom, respectively.
Moreover, making the investigation of the ground state more challenging, the apparent structure along Si step-edge chains in scanning tunneling microscopy (STM) changes into another $\times$2 ordering below about 60 K \cite{Polei2013,Polei2014}.
The atomic structure of this ordering is not clear at all but was suggested to be induced by the tunneling current \cite{Brihuega2005,Polei2013}.
This means that STM cannot probe the pristine ground state at low temperature.

%%%%%%%%%%%%%%%%%%%%%%%%%% Transition %%%%%%%%%%%%%%%%%%%%%%%%%%
In the present work, we report an atomic resolution noncontact (nc) atomic force microscopy (AFM) study on the Si(553)-Au surface together with simultaneous STM measurements.
These combined measurements provide effective decoupling of the possible excitation source (tunneling current) and the local structural probe (AFM).
The unperturbed step-edge Si structure at 4.3 K is observed to possess a trimer ground state with every third atom distorted downward and with an insulating property.
This clearly rules out the spin ordering model \cite{Erwin2010,Hafke2016} and supports strongly the charge/structure-modulated model \cite{Braun2018}.
The present model corresponds to a CDW ground state with a triple degeneracy or a $Z_{3}$ topology for the Si step-edge chain \cite{Ahn2005,Snijders2006,Liu2017}, whose electronic states are unexpectedly decoupled from those of neighboring Au chains.
The tunneling current injection is found to induce an apparently undistorted structure in AFM but a buckled dimer structure in STM.
This structure is explained by the fluctuation between three translationally degenerate trimer structures with their translational barrier reduced substantially by the tunneling current.
The present work thus solves a long standing puzzle on the structure of this model 1D electronic system.
Beyond that, this work discloses the topological nature of the present system and introduces it as a platform for manipulating topological phases and topological excitations such as $Z_{3}$ solitons.
The combination of atomic resolution STM and AFM can be widely applied to decouple not only the electronic modulation from atomic structure but also the electronic local excitation source for manipulation from the structural probe.

%%%%%%%%%%%%%%%%%%%%%%%%%% Method %%%%%%%%%%%%%%%%%%%%%%%%%%

%%%%%%%%%%%%%%%%%%%%%%%%%% Result %%%%%%%%%%%%%%%%%%%%%%%%%%

%%%%%%%%%%%%%%%%%%%%%%%%% Insert Figures %%%%%%%%%%%%%%%%%%%%%%%%%

\begin{figure}[!tb] %[t], [b], ...
\includegraphics[width=8.6cm]{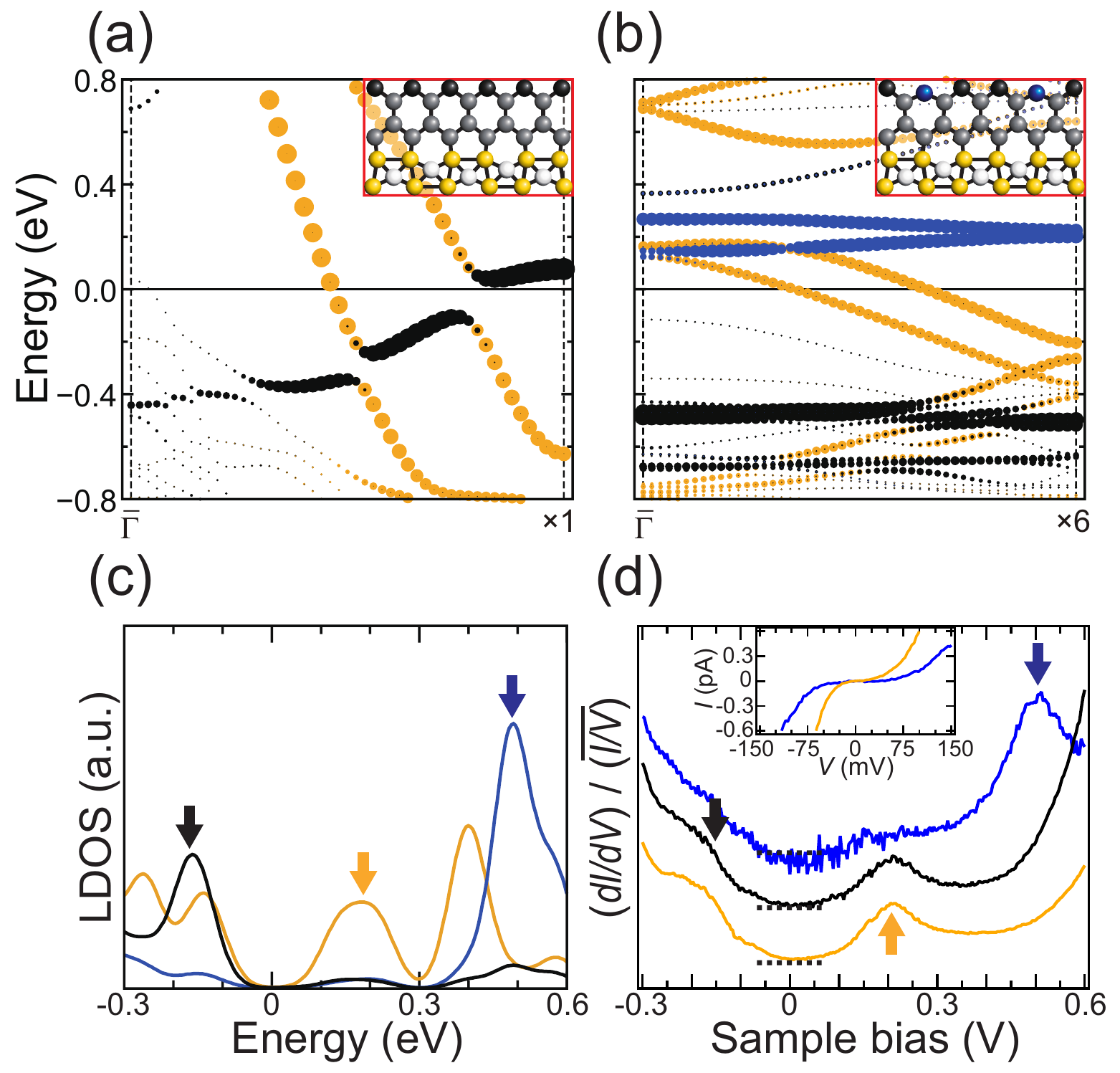}
\caption{\label{f2} Calculated band structures of (a) undistorted and (b) distorted surfaces along the step-edge direction. The insets show the schematics of the structure models, respectively.} (c) Calculated and (d) measured local density of states (LDOS) spectra on the step-edge Si chain (blue and black for distorted and undistorted atoms, respectively) and on the terrace Au chain (yellow). For the calculated spectra, energy is shifted for +0.3 eV for comparison. The $dI/dV$ spectra are normalized conductance (the method described in \cite{Feenstra1994}), and were measured at 50 K to prevent the unintended current-induced transition. Curves are vertically offset for comparison (dashed lines indicate zeros). Five prominent spectral features are specified on the curves (blue, black, and yellow arrows). Inset: $I-V$ curves recorded from $\times$3 step-edge Si (blue) and terrace Au (yellow) chains. In comparison with the calculated LDOS, one has to consider a roughly parabolic background in the experiment centered at zero bias.

\end{figure}

\begin{figure*}[!tb] %[t], [b], ...
\includegraphics[width=15.3cm]{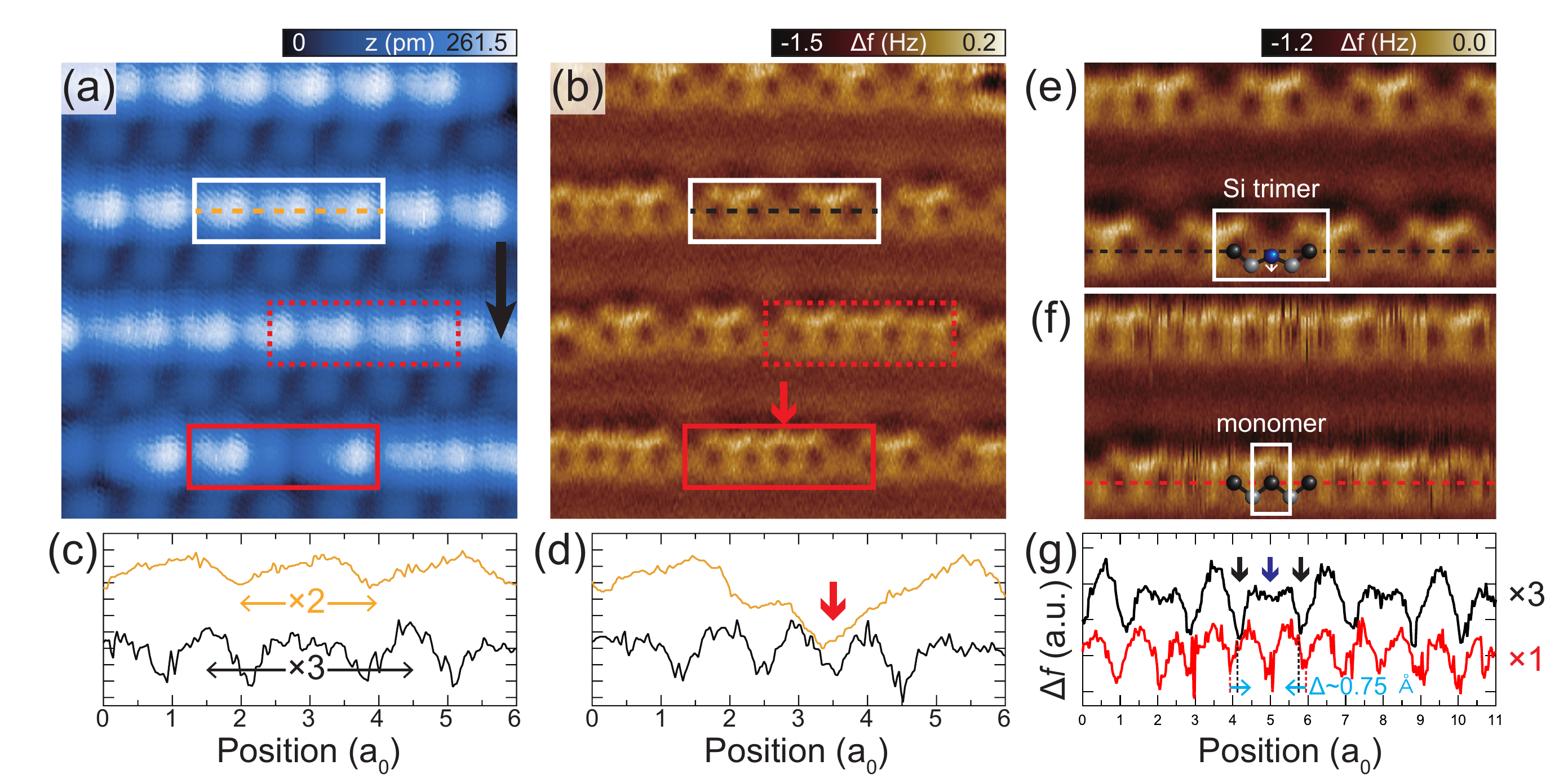}
\caption{\label{f3} Topographic (a) STM and (b) AFM images (5.6$\times$5.6 nm$^2$) of Si(553)-Au at 4.3 K. The images are taken at +1.0 (STM, with $I$=10 pA}) and zero (AFM) sample biases. Black arrow indicates the uphill direction of steps. Comparison of STM (yellow) and AFM (black) line profiles taken across the (c) Si step-edge wire (white rectangle) and (d) static adsorbate impurity \cite{Shin2012} (solid red). Bias-dependent AFM images measured at (e) zero and (f) +0.5 V sample biases. (g) $\Delta f$ line profiles taken across the Si step-edge chain at zero (black) and +0.5 V (red) which exhibit two different surface states of trimer and monomer, respectively. Down- (un)distorted atom of the trimer structure is represented by blue (black) sphere.

\end{figure*}

\begin{figure*}[!tb] %[t], [b], ...
\includegraphics[width=15.3cm]{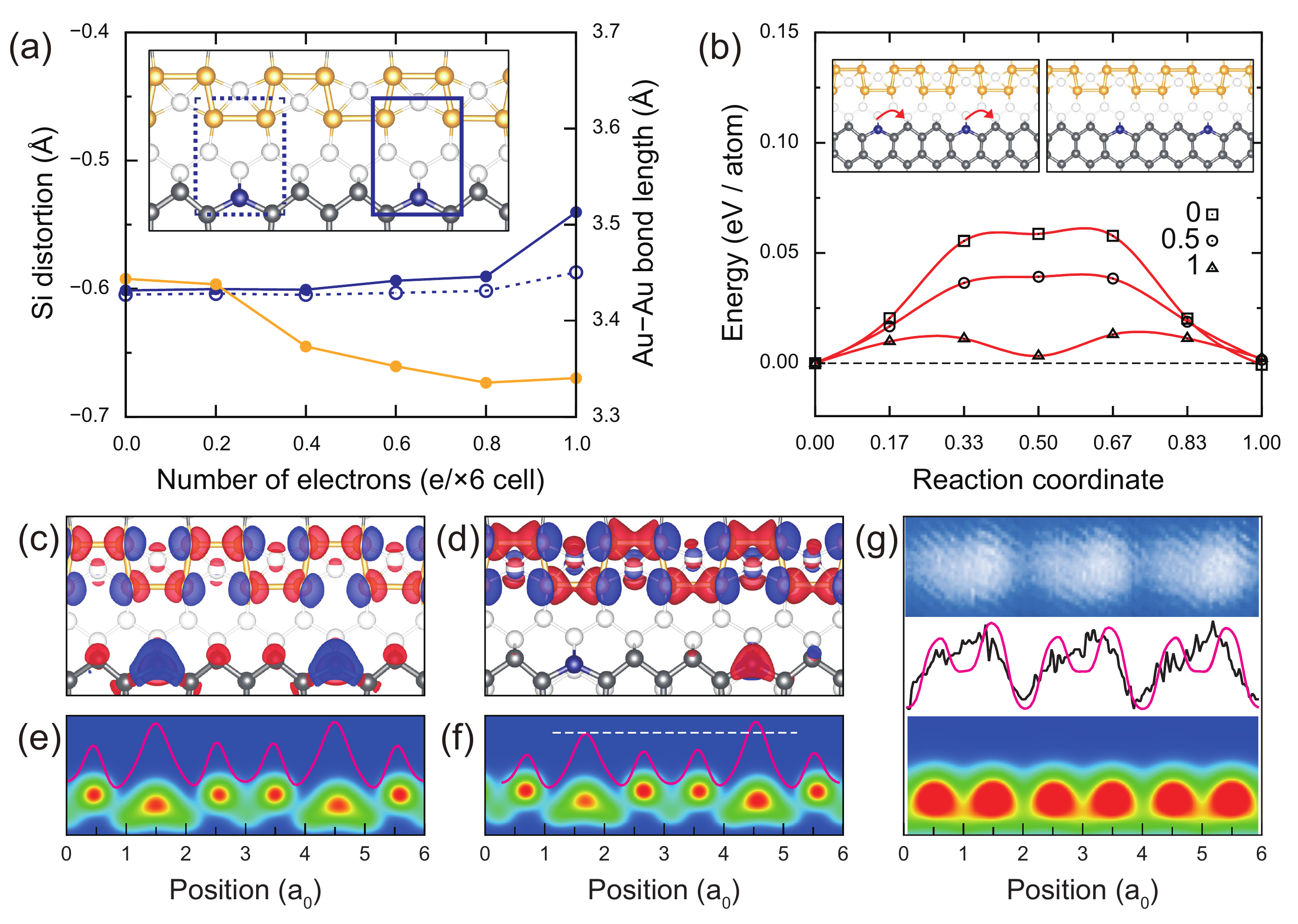}
\caption{\label{f4} Electron doping effect on a CDW ground state of the Si(553)-Au surface. (a) Structural distortions of the Si step-edge chain (blue) and the Au terrace chain (yellow) as a function of electron doping. Inset: solid and dashed rectangles indicate the distorted Si atoms placed close to the long and the short Au-Au bonds, respectively. (b) Translational energy barriers between the degenerate ground states for different electron doping levels. Inset: initial and final states for a diffusion process. (c,d) Iso-surface of total charge difference ($\Delta \rho$). (c) $\Delta \rho$=$\rho_{CDW}$-$\rho_{\times 1}$ ($7\times10^{-3}$ e{\AA}$^{-3}$) for undoped surface ($\rho_{0.0}$). (d) $\Delta \rho$=$\rho_{1.0}$-$\rho_{0.0}$ ($2\times10^{-3}$ e{\AA}$^{-3}$). Red (blue) color indicates positive (negative) charges. Simulated STM images ($V$=+1.0 V and $z$=2 {\AA}) for (e) undoped and (f) doped (1.0e) $\times$6 supercells. Magenta lines indicate $z$ profiles along the step-edge chain. (g) Comparison between the STM image at 4.3 K (top) and the simulated one ($V$=+1.0 V and $z$=4 {\AA}) averaged for all three degenerate ground states (bottom).
}
\end{figure*}

%%%%%%%%%%%%%%%%%%%%%%%%%Fig1%%%%%%%%%%%%%%%%%%%%%%%%%
\noindent{\fontfamily{phv}\selectfont \textbf{RESULTS AND DISCUSSION}}

% Atomic structure
Figures \ref{f1}(b) and \ref{f1}(e) show the well known empty- and filled-state STM images taken at 50 K along a step edge of the Si(553)-Au surface, respectively.
The brightest protrusions in the STM images are due to Si step-edge atoms [Fig. \ref{f1}(f)].
At this temperature, the perturbation by the tunneling current \cite{Polei2013,Polei2014} is not substantial and the step-edge exhibits the well known $\times$3 superstructure \cite{Erwin2010,Hafke2016,Braun2018}.
On the other hand, the atomic chains on narrow terraces show a $\times$2 superstructure with a much reduced contrast [indicated by a yellow arrow in Fig. \ref{f1}(f)].
These results are fully consistent with many previous STM works \cite{Ahn2005,Shin2012,Polei2013,Polei2014,Aulbach2013,Aulbach2017,Hafke2016}.
Note also the strong bias dependence of the STM images [Figs. \ref{f1}(b)--\ref{f1}(e)], which indicates a substantial electronic effect for these images such as the spin splitting \cite{Erwin2010,Hafke2016} or the rehybridization of dangling bonds \cite{Braun2018}.

In stark contrast, a $\Delta f$ image of nc-AFM can reveal directly the structural distortion, if any, since it sensitively detects the relative distance between the tip and individual surface atoms.
The corresponding image of Fig. \ref{f1}(f) exhibits strong negative $\Delta f$ (the dark contrast in the image), which indicates surface atoms closer to the tip.
This image tells immediately that the step edge is structurally distorted with a significant height difference.
Since the height (contrast) of the two (black) atoms are similar to those of the terrace atoms, we can conclude that every third (blue) atoms of the step edge has a downward distortion.
We also observe the alternating bright and dark contrast in a $\times$2 period along the Au chains within the narrow terraces between Si step edges in the AFM image (pointed by the yellow arrow), which is consistent with what observed in the STM images.
Since the Au atomic positions would be represented by dark contrast, this image indicates a dimerlike paring of Au atoms.
The former is crucial for the debate on the surface structure since the recent structure models share the double Au row structure with a $\times$2 distortion along the terrace but differ substantially in the structural distortion on the Si step edge \cite{Krawiec2010,Erwin2010,Hafke2016,Braun2018}.
Namely, the previous calculations and the present one indicate that the structural distortion of the antiferromagnetic spin chain model is negligible in clear contrast to the present AFM observation even though this model gives a strong electronic modulation in STM similar to that of the rehybridized model.
On the other hand, the strong step-edge distortion revealed in the nc-AFM image is consistent with the rehybridized step-edge chain model [Fig. \ref{f1}(f)] \cite{Braun2018,Erwin2010,Hafke2016}.
Furthermore, the atomic origin of the STM protrusions can be disclosed unambiguously by the high resolution $\Delta f$ image of the same structure taken simultaneously with STM.
The strong empty-state protrusion is localized on the down-distorted Si atom at the step edge and the double protrusions in the filled-state on the undistorted Si atoms.
This is fully consistent with the result of the DFT calculations based on the reconstruction of step-edge Si atoms within the rehybridization model \cite{Braun2018}.
This reconstruction, the downward distortion on every third atoms, splits the $\textit{sp$_3$}$ dangling bond into $\textit{sp$_2$}$ and $\textit{p}$ orbitals in filled- and empty-states, respectively, which explains the corresponding STM contrast \cite{Braun2018}.

%%%%%%%%%%%%%%%%%%%%%%%%%Fig2%%%%%%%%%%%%%%%%%%%%%%%%%
% Electronic states
The reconstruction of step-edge atoms would be reflected in surface electronic states.
Figures \ref{f2}(a) and \ref{f2}(b) show the band structures calculated along step edges.
In the undistorted structure [Fig. \ref{f2}(a)], the Au chains have two strongly metallic bands coming from the double Au rows in consistency with the previous angle-resolved photoemission measurements \cite{Ahn2005,Yeom2014}.
Note that the spin-orbit coupling is not included here, which explains small extra splittings observed in these Au 6$\textit{s}$ bands \cite{Yeom2014}.
The silicon dangling bonds along step edges also form a metallic band.
This is naturally expected from the partially filled nature of dangling bonds.
The Au and Si bands are hybridized, and the latter receives electrons from the substrate to be filled over the half filling.
As mentioned above, the step-edge reconstruction splits the Si band into the empty $\textit{p}$ band just above the Fermi level and the fully filled $\textit{sp$_2$}$ band [blue and black circles in Fig. \ref{f2}(b), respectively].
These states are rather well localized on the distorted and undistorted Si atoms, respectively.
The $\times2$ distortion on the Au chains also makes a band gap, but it is located above the Fermi level around +0.4 eV.
Thus, the Au chains keep its metallic property but the Si chains become insulating with a gap larger than 0.6 eV.
These calculated results explain the main features of the measured STS ($dI/dV$) spectra [Fig. \ref{f2}(d)].
Most importantly, the density of states for Si chains of the reconstructed surface exhibit a substantial splitting of the main spectra with the strong localization of the empty-state feature (+0.50 eV) on the distorted atom (the blue arrow) and the filled-state (-0.16 eV) on the undistorted atoms (the black arrow).
The empty state closer to the Fermi level at +0.2 eV is due to the electronic state related to Au chains.
This state appears also in the spectra of the Si chain atoms, which seems to indicate the limited spatial resolution of this particular STS measurement.
The insulating property of the Si chains can be clearly deduced from the STS measurements in contrast to the Au chains [inset of Fig. \ref{f2}(d) and see Fig. S1 in the supplementary information for more the detailed gap measurements].
The major discrepancy between the calculation and the measurement is the rigid shift of the spectra by 0.3 eV [Figs. \ref{f2}(c) and \ref{f2}(d)], which obviously indicates a less charge transfer from the Si substrate than calculated.
A smaller charge transfer (or a hole doping) puts the Fermi level of the Si step-edge chain well within its band gap.
Within this situation, namely with the large band gap formation and the shift of the Fermi level for the Si dangling-bond band, the insulating property of the Si step-edge chain becomes robust, and the coupling of the bands of Si and Au chains becomes much less significant.
In our additional calculations, we prove that the mild hole doping indeed leads to a rigid shift of the Si states and is favorable for the $\times$3 CDW formation on the step-edge Si [Fig. S2 in the supplementary information].
In contrast, the electron doping suppresses the the $\times$3 structure, which will be discussed below.

The distorted and insulating Si step-edge chains straightforwardly indicate a CDW insulator although its mechanism is largely different from most of known CDW systems.
The mechanism is not fully consistent with the Fermi surface nesting of a dispersive band but is much closer to the Jahn-Teller distortion.
However, the present band gap is not a bonding-antibonding splitting of the Jahn-Teller effect but an orbital rehybridization or the orbital ordering of \textit{p} and \textit{sp$_{2}$} orbitals.
Since the distortion has a $\times$3 periodicity, there exist three degenerate ground states.
This corresponds to a realization of a $Z_{3}$ topological insulator, whose material realization has not been reported to our best knowledge.
Topological excitations or solitons of this system are expected to have fractional charges of (2/3)e and (4/3)e \cite{Su1979,Su1980,Liu2017}.
The soliton excitation of the present system was suggested in an early STM work for local defects \cite{Snijders2006} and in the recent electron diffraction study for the thermally induced disordering \cite{Shin2012,Hafke2020,Braun2020}, and was identified experimentally very recently \cite{Park2021}.
The present structural information provides a solid ground for the existence of such exotic topological solitons.

%%%%%%%%%%%%%%%%%%%%%%%%%Fig3%%%%%%%%%%%%%%%%%%%%%%%%%
As mentioned above, the STM images at lower temperature were reported to exhibit different superstructures, developing gradually a new $\times$2 superstructure \cite{Polei2013,Polei2014}.
Figure \ref{f3}(a) shows an empty-state STM image recorded at 4.3 K where the effect of tunneling current was reported to dominate.
It is apparent that both chains have a 2$a_{0}$ periodicity with Si step-edge atoms in apparent up-and-down buckling contrast.
However, at the same temperature and position, the nc-AFM image shows mainly the $\times$3 superstructure [Figs. \ref{f3}(b) and \ref{f3}(c)], which is consistent with the measurements at 50 K.
Thus, we can assure that the $\times$2 structure on the step edge is indeed induced by tunneling current as suggested in the previous works \cite{Polei2013,Polei2014}.
The effect of the tunneling current can further be corroborated by the nc-AFM measurement with the tunneling current on; while the AFM images shown so far are obtained with the null tip bias [Fig. \ref{f3}(e)], we can also obtain the same quality image with the tunneling current (bias of 0.5 V) [Fig. \ref{f3}(f)].
Figures \ref{f3}(e)--\ref{f3}(g) unambiguously indicate that the trimer reconstruction on the step-edge is completely relieved by the tunneling current and the current-induced transition was observed to be reversible [Fig. S3].
This confirms that the ground state reconstruction is based on the $\times$3 and $\times$2 superstructures along the Si step-edge and Au terrace chains, respectively.
In fact, the apparently undistorted structure is observed locally in the AFM image without the tunneling current in Fig. \ref{f3}(b) (dashed rectangle), which is thought to be due to the presence of local defects.

%We further suggest that the buckling in the STM image has an electronic origin as it is not reflected in the AFM contrast.
%The electronic modulation or marginal structural distortion of a 2$a_{0}$ periodicity is naturally expected since the neighboring Au chains are paired structurally to result in a periodic strain field and their electronic states are partially hybridized with the Si dangling bonds at step-edges \cite{Braun2018,Braun2020}.

%%%%%%%%%%%%%%%%%%%%%%%%%Fig4%%%%%%%%%%%%%%%%%%%%%%%%%
The origin of this additional structure with a weak buckling in the STM image can be explained by the electron doping effect within our theoretical calculations.
The transient electron doping by tunneling current was already proposed to explain the tunneling-induced structure within the antiferromagnetic spin chain model \cite{Polei2013,Polei2014}.
This model suggested that a 2$a_{0}$-periodic antiferromagnetic spin ordering is favored over the 3$a_{0}$ one when the system is doped.
This model however cannot explain the much reduced STM contrast in the $\times$2 superstructure as compared with that of $\times$3, since the $\times$2 and $\times$3 antiferromagnetic orderings of spins produce the same amount of exchange splitting and electronic modulation.
In Fig. \ref{f4}, the structural and electronic properties of the surface were examined for various electron doping levels by using the present rehybridized model \cite{Braun2018}.
The additional electron doping enhances the Au-Au bond length to form a stronger $\times$2 distortion of the Au chain [Fig. \ref{f4}(a)], which is consistent with the previous theoretical study \cite{Mamiyev2018}.
In stark contrast, the doping relieves the distortion of Si step-edge atoms to suppress the $\times$3 CDW order, especially after the saturation of the Au dimerization around 0.8e doping.
This indicates that doped electrons fill the metallic Au bands [Figs. \ref{f2}(b) and \ref{f4}(c)] at the initial stage and, at a higher doping level, accumulate on the unoccupied part of the distorted Si bands to destabilize the distorted structure [Figs. \ref{f4}(c) and \ref{f4}(d)].
As a result, the energy gain of the $\times$3 CDW structure is significantly reduced together with the transition energy barrier between three translationally degenerate $\times$3 structures [Fig. \ref{f4}(b)].
Since three degenerate ground states compete each other with a small energy barrier of 13 meV, at a finite temperature, the doped system would exhibit a dynamic fluctuation between them.
The averaged STM image over three degenerate states matches excellently with the weakly buckled $\times$2 structure observed by STM at low temperature [Fig. \ref{f4}(g)].
The weak buckling is the effect of the strong Au dimerization, which affects the translational potential energy between the degenerate states (see the supplementary information Fig. S4.

%%%%%%%%%%%%%%%%%%%%%%%%%% Conclusion %%%%%%%%%%%%%%%%%%%%%%%%%%
\noindent{\fontfamily{phv}\selectfont \textbf{CONCLUSIONS}}

The atomic and electronic structures of the Si(553)-Au chain system are investigated by the atomic resolution nc-AFM together with simultaneous STM/STS measurements.
These measurements unambiguously reveal the strong trimer structural distortion and a substantial band gap along the step-edge Si chains at low temperature, which corresponds to a CDW ground state and a $Z_{3}$ topological insulator.
The transient electron doping by the tunneling current is shown to reduce the stability of the CDW phase to induce a dynamically fluctuating CDW state at very low temperature.
Further studies on topological properties and topological excitations in this system are thus highly promising.
In particular, the exploitation of the $Z_{3}$ topological solitons in the present system would provide opportunity for the topologically-protected multilevel information processing.
The combination of atomic resolution STM and AFM is shown to be useful to decouple the electronic local excitation source for manipulation from the structural probe.

\noindent{\fontfamily{phv}\selectfont \textbf{METHODS}}

Atomically-resolved imaging and spectroscopy of the Si(553)-Au surface were performed by simultaneous high-resolution nc-AFM and STM measurements under ultra high vacuum at two different temperatures of 4.3 and 50 K using a commercial low-temperature microscope (SPECS GmbH) with force-current detection capabilities.
A tungsten tip mounted on a tuning fork was used for both AFM and STM measurements.
AFM images were obtained by measuring the frequency shift ($\Delta f$) of the tuning fork at constant height mode.
For scanning tunneling spectroscopy (STS) measurements, current-voltage ($I$-$V$) curves were recorded with the lock-in technique using a modulation amplitude of 10 mV.
The clean Si(553) surface was prepared by repeated flash heating to 1520 K.
Gold was evaporated onto the clean Si(553) surface held at 923 K.
At a Au coverage of around 0.5 monolayer, well-ordered Si(553)-Au surfaces were reproducibly fabricated \cite{Shin2012,Aulbach2017}.
DFT calculations were performed using the Vienna \textit{ab initio} simulation package \cite{Kresse1996} within the generalized-gradient approximation (GGA) using the revised Perdew-Burke-Ernzerhof (PBEsol) functional \cite{Perdew2008}.
The Si(553)-Au surface is modeled within periodic supercells with six bulklike Si layers and a vacuum spacing of about 12.8  {\AA}.
The bottom of the slab was passivated by H atoms.
We used a plane-wave basis with an energy cutoff of 312 eV and 5$\times$2$\times$1 k-point mesh.
All atoms but the bottom two Si layers were relaxed until the residual force components were within 0.03 eV/{\AA}.
%For the tight-binding model calculations, we used the PythTB package by Coh and Vanderbilt \cite{PythTB}.

\noindent{\fontfamily{phv}\selectfont \textbf{ASSOCIATED CONTENTS}}

\noindent{\fontfamily{phv}\selectfont \textbf{Supporting Information}}

The Supporting Information is available free of charge at https://pubs.acs.org/.

In the supporting information, we provide further discussion on the quantitative estimation of the band gap (Fig. S1), the stability of the CDW phase upon doping (Fig. S2), the reversibility of current-induced transition (Fig. S3), and the STM image simulation for the current(doping)-induced structure (Fig. S4).

\noindent{\fontfamily{phv}\selectfont \textbf{AUTHOR INFORMATION}}

\noindent{\fontfamily{phv}\selectfont \textbf{Author contributions}}
E.D. performed STM and AFM experiments and analyzed data. J.W.P. performed DFT and model calculations. O.S. and P.V. helped AFM experiments. H.W.Y. conceived the project idea, directed experiments/calculations. E.D., J.W.P., and H.W.Y. wrote the manuscript. E.D. and J.W.P. equally contributed to this work.

%%%%%%%%%%%%%%%%%%%%%%%%%% Acknowledgement %%%%%%%%%%%%%%%%%%%%%%%%%%%%%
\noindent{\fontfamily{phv}\selectfont \textbf{ACKNOWLEDGEMENTS}}

This work was supported by the Institute for Basic Science (Grant No. IBS-R014-D1) and by the Czech Science Foundation grants (14-374527G). We would like to acknowledge fruitful discussions on AFM experiments with Martin Svec.

%Bibtex part%
\bibliography{Si553-Au_str} %bibliography file name%

\end{document}